\documentclass[sigconf]{acmart}
\usepackage{verbatim}
\usepackage{array}
\usepackage{makecell}
\usepackage{float}
\usepackage{bm}
\usepackage{graphicx}
\usepackage{threeparttable}

\AtBeginDocument{%
  \providecommand\BibTeX{{%
    \normalfont B\kern-0.5em{\scshape i\kern-0.25em b}\kern-0.8em\TeX}}}

\setcopyright{acmcopyright}
\copyrightyear{2020}
\acmYear{2020}

\acmConference[DLP-KDD '20]{DLP-KDD '20: 2nd Workshop on Deep Learning Practice for High-Dimensional Sparse Data with KDD 2020}{Aug 24, 2020}{San Degio, CA}
\acmBooktitle{ 2nd Workshop on Deep Learning Practice for High-Dimensional Sparse Data with KDD 2020, Aug 24, 2020, San Degio, CA}
\acmPrice{15.00}

\begin{document}

\title{COLD: Towards the Next Generation of Pre-Ranking System}



\author{Zhe Wang, Liqin Zhao, Biye Jiang, Guorui Zhou, Xiaoqiang Zhu, Kun Gai}
\affiliation{%
  \institution{Alibaba Group}
   \city{Beijing}
   \country{P.R.China}
}
\email{{wz143459, liqin.zlq, biye.jby, guorui.xgr, xiaoqiang.zxq, jingshi.gk}@alibaba-inc.com}

\renewcommand{\shortauthors}{Wang and Zhao, et al.}


\begin{abstract}

Multi-stage cascade architecture exists widely in many industrial systems such as recommender systems and online advertising, which often consists of sequential modules including matching, pre-ranking, ranking, etc.    
For a long time, it is believed pre-ranking is just a simplified version of the ranking module, considering the larger size of the candidate set to be ranked. Thus, efforts are made mostly on simplifying ranking model to handle the explosion of computing power for online inference. For example, SOTA pre-ranking solution of display advertising systems is to restrict the pre-ranking model to follow a vector-product based deep learning architecture: user-wise and ad-wise vectors are pre-calculated in an offline manner with no user-ad cross features, then the inner product of the two vectors is calculated online to obtain the pre-ranking score. Obviously, this kind of model restriction results in suboptimal performance.  

In this paper, we rethink the challenge of the pre-ranking system from an algorithm-system co-design view. Instead of saving computing power with restriction of model architecture which causes loss of model performance, here we design a new pre-ranking system by joint optimization of both the pre-ranking model and the computing power it costs.  We name it COLD (\textbf{C}omputing power cost-aware \textbf{O}nline and \textbf{L}ightweight \textbf{D}eep pre-ranking system). 
COLD beats SOTA in three folds: 
(i) an arbitrary deep model with cross features can be applied in COLD under a constraint of controllable computing power cost.
(ii) computing power cost is explicitly reduced by applying optimization tricks for inference acceleration. This further brings space for COLD to apply more complex deep models to reach better performance.   
(iii) COLD model works in an online learning and severing manner, bringing it excellent ability to handle the challenge of the data distribution shift. Meanwhile, the fully online pre-ranking system of COLD provides us with a flexible infrastructure that supports efficient new model developing and online A/B testing.  
Since 2019, COLD has been deployed in almost all products involving the pre-ranking module in the display advertising system in Alibaba, bringing significant improvements.

\end{abstract}





\maketitle

\begin{figure}[th]
    \begin{center}  
      \includegraphics[width=0.4\textwidth] {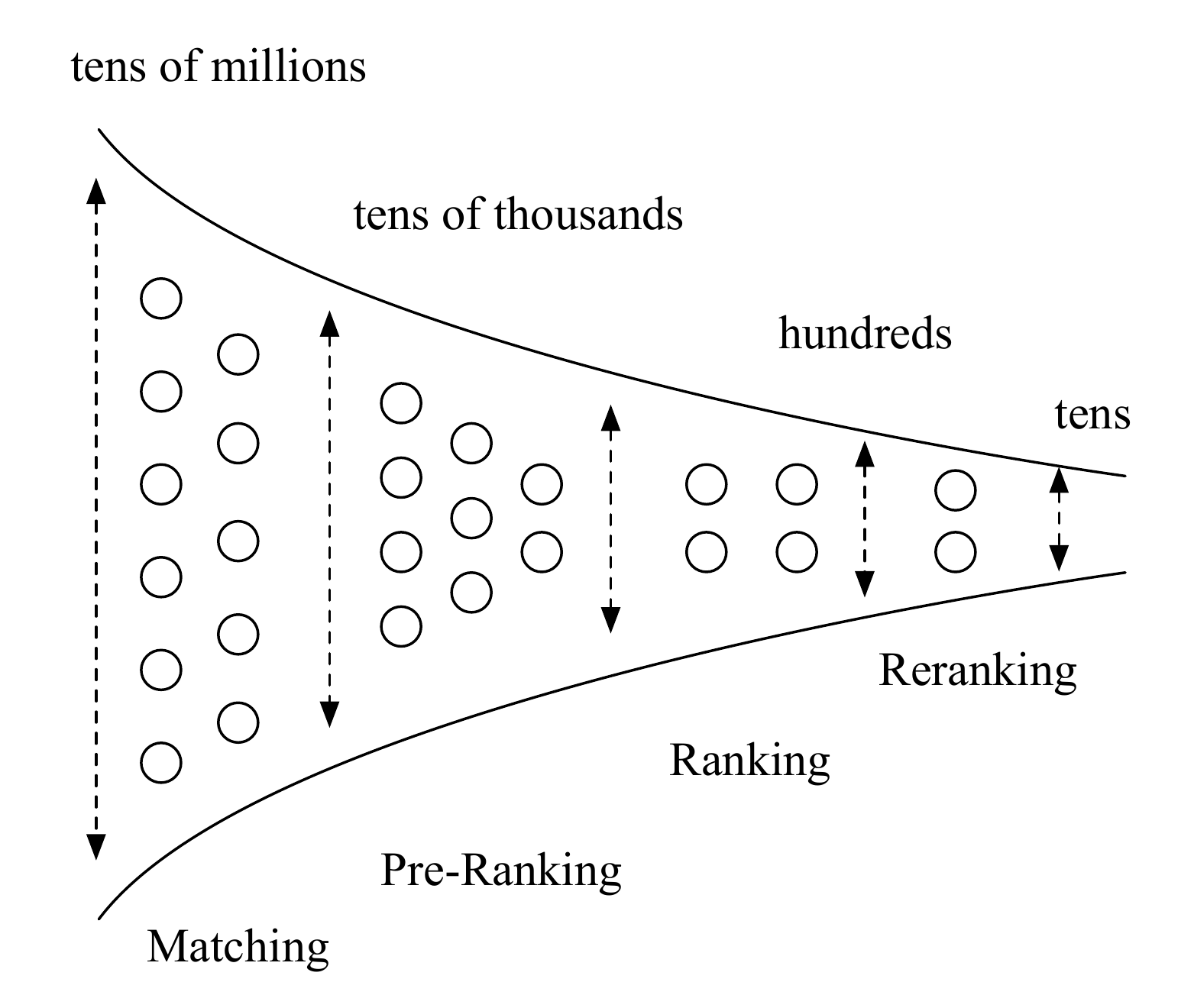} \\
      \caption{Illustration of cascade architecture for industrial information retrieval system.}
      \label{intro:cascade}
    \end{center}
  \end{figure}

\section{Introduction}

\begin{figure*}[pt]
    \begin{center}
\centering
      \includegraphics[width=0.9\textwidth] {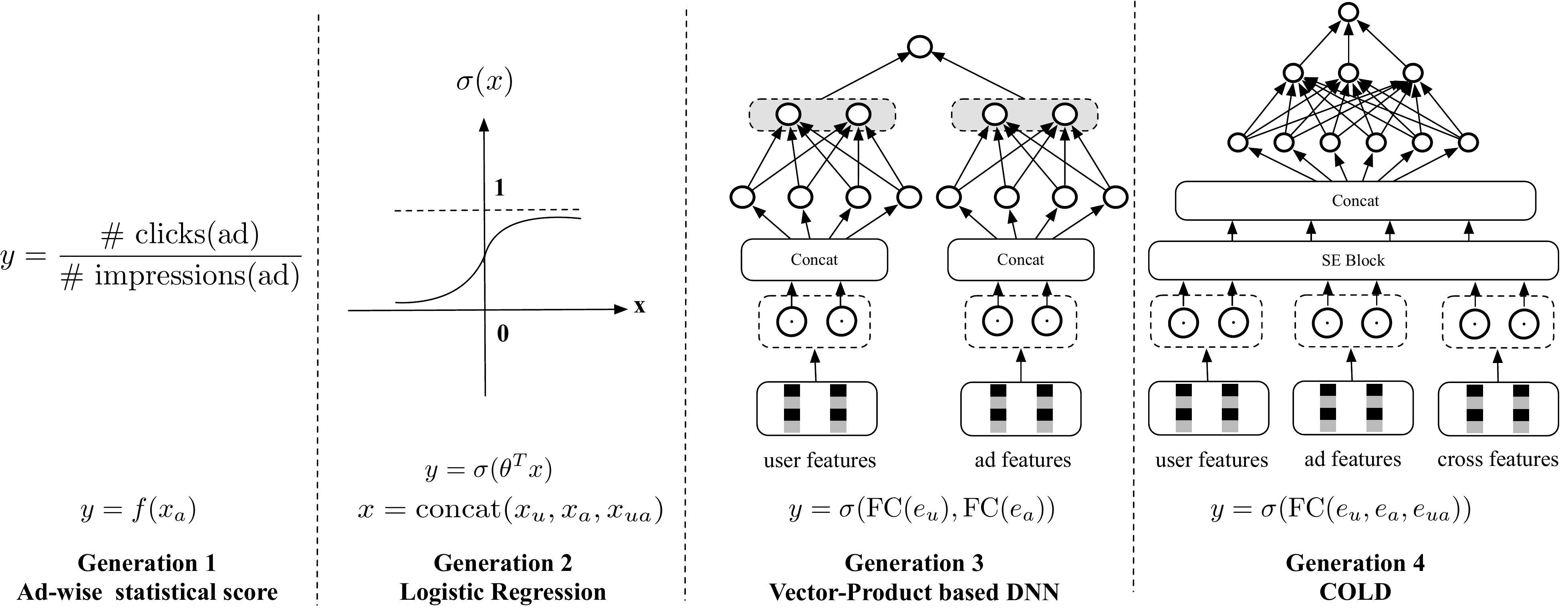} 
      \caption{The development history of the pre-ranking system from model view. $x_u,x_a,x_{ua}$ are the raw features of user, ad and cross. $e_u,e_a,e_{ua}$ are the embeddings of user, ad, and cross features. The first generation is the non-personalized ad-wise statistical model.  LR (Logistic Regression) model is the second generation which is a lightweight version of the large scale ranking model in the age of shallow machine learning \cite{cascaderanking}. It can be deployed in online learning and serving manner. Vector-product based deep learning architecture is the third generation and current state-of-the-art pre-ranking model. It significantly boosts the model performance over the previous generation. COLD is our proposed new generation of the pre-ranking model. }
          \label{cmp}
    \end{center}
  \end{figure*}
    
Users have been struggling with information overload due to the rapid growth of internet services these years. Search engine, recommender systems, and online advertising have become foundational information retrieval applications which server billions of users every day. Most of these systems \cite{youtube:recommend,cascaderanking,zhou2018deep,Wu:2018uo,zhou2019dien,Chen:2019tc,pi2019deep,Fan:2019ge} follow a multi-stage cascade architecture, that is, candidates are extracted by sequential modules such as matching, pre-ranking, ranking, and reranking, etc. Figure \ref{intro:cascade} gives an brief illustration.  

There have been numerous papers discussing how to build an effective and efficient ranking system \cite{youtube:recommend,gai2017learning,Grbovic:2018bw,zhou2018deep,zhou2019dien,Wu:2018uo,pi2019deep,Chen:2019tc,Fan:2019ge,Lyu:2020ue}. However, very few works pay attention to the pre-ranking system \cite{cascaderanking,Wu:2018uo}. For simplicity, in the rest of the paper, we limit ourselves to discuss the design of the pre-ranking system in the display advertising system. Techniques discussed here can be easily applied in recommender systems, search engines, etc.

For a long time, it is believed that pre-ranking is just a simplified version of the ranking system, considering the computing power cost challenge of online serving with a larger size of the candidate set to be ranked. Take the display advertising system in Alibaba as an example. Traditionally, the size of the candidate set to be scored for the pre-ranking system scales up to tens of thousands, which turns to be hundreds in the subsequent ranking system. On the other side, both ranking and pre-ranking systems have strict latency limit, e.g., $10 \sim 20$ milliseconds. In this situation, the pre-ranking system is often designed as a lightweight ranking system by simplifying the ranking model to handle the explosion of computing power for online inference. 
    
\subsection{Brief Introduction of the Development History of the Pre-Ranking System}   
Looking back at the development history of the pre-ranking system in industry, we can simply categorize it into four generations from the model view, as shown in figure \ref{cmp}.   
The first generation is the non-personalized ad-wise statistical score. It calculates the pre-rank score by averaging out the recent CTR (Click-Through Rate) of each ad. The score can be updated with high frequency. 
LR (Logistic Regression) model is the second generation which is a lightweight version of the large scale ranking model in the age of shallow machine learning. It can be deployed in the online learning and serving manner \cite{McMahan:2013cq}.
Vector-product based deep learning model \cite{youtube:recommend} is the third generation and current state-of-the-art pre-ranking model. In this method, user-wise and ad-wise embedding vectors are pre-calculated separately in an offline manner with no user-ad cross features, then the inner product of the two vectors is calculated online to obtain the pre-ranking score. Although vector-product based DNN has significantly boosted the model performance of the first two generations, it still suffers from two challenges which leave space for further improvement: (i) \textbf{Model expression ability}. As shown in \cite{Zhu:2018gi}, the expression ability of the model is limited by restricting deep models with vector-product form; (ii) \textbf{Model update frequency}. 
The embedding vectors of the vector-product based DNN need to be pre-calculated offline and then loaded into the memory of the server for online calculation.
It means the vector-product based DNN model can only be updated in a low-frequency manner, which make it hard to adapt to the newest data distribution shift, especially when the data changes dramatically (e.g., Double 11 event in China).  


To summarize, all the above mentioned three generations of the pre-ranking system follow the same paradigm: computing power is treated as a constant constraint, under which the pre-ranking model corresponding with the training and serving system is developed. That is, the design of the model and the optimization of the computing power is decoupled, which usually leads to a simplification of the model to fit the requirement of computing power. This results in suboptimal performance.   
      
\subsection{COLD: New Generation of Pre-Ranking System}      
In this paper, we rethink the challenge of the pre-ranking system from an algorithm-system co-design view. Instead of saving computing power with restriction of model architecture which limits the performance, here we design a new pre-ranking system by jointly optimizing both the pre-rank model and the computing power it costs. We name it COLD (\textbf{C}omputing power cost-aware \textbf{O}nline and \textbf{L}ightweight \textbf{D}eep pre-ranking system), as illustrated in  Figure \ref{cmp}. 
We treat COLD as the fourth generation of the pre-ranking system. COLD takes into consideration of both model design and system design. Computing power cost in COLD is also a variable that can be optimized jointly with model performance. In other words, COLD is a flexible pre-ranking system that the trade-off between model performance and computing power cost is controllable. 

Key features of COLD are summarized as follows:
\begin{itemize}
    \item Arbitrary deep model with cross features can be applied in COLD under a constraint of controllable computing power cost. In our real system, COLD model is a seven-layered fully connected deep neural network with SE (  Squeeze-and-Excitation) block \cite{hu2018squeeze}. SE block benefits us to conduct feature group selection to get a lightweight version from a complex ranking model. This selection is executed by taking into consideration of both model performance and computing power cost. That is, computing power cost for COLD model is controllable. 
    \item Computing power cost is explicitly reduced by applying optimization tricks such as parallel computation and semi-precision calculation for inference acceleration. This further brings space for COLD to apply more complex deep models to reach better performance.
    \item COLD model works in an online learning and severing manner, bringing system excellent ability to handle the challenge of data distribution shift.  The fully online pre-ranking system of COLD provides us with a flexible infrastructure that supports new model developing and fast online A/B testing, which is also the best system practice currently that ranking system owns.  
\end{itemize}

\begin{figure}[t]
    \begin{center}  
      \includegraphics[width=0.4\textwidth] {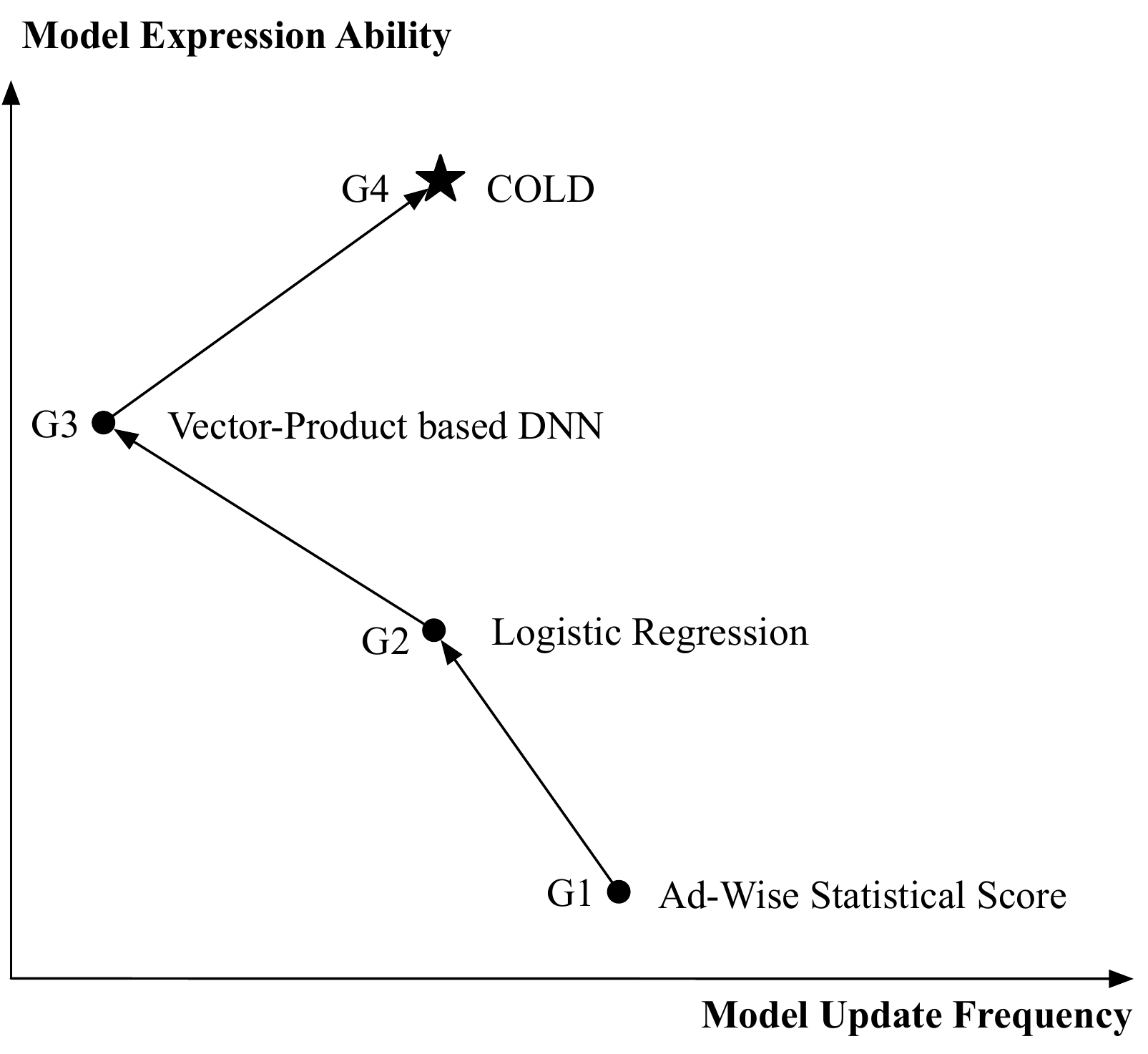} \\
      \caption{Model expression ability v.s. update frequency of the four generations of ranking system.}
      \label{intro:trend}
    \end{center}
  \end{figure}
  
Figure \ref{intro:trend} gives a comparison of all the four generations of ranking system w.r.t. model expression ability and update frequency. COLD achieves the best tradeoff. Since 2019, COLD has been deployed in almost all products involving the pre-ranking module in the display advertising system in Alibaba, serving hundreds of millions of users with high concurrent requests every day. Compared with vector-product based DNN, our latest online version of the pre-ranking model, COLD brings us more than 6\% RPM improvement, which is a significant improvement for the business.    

The rest of the paper is organized as follows: section 2 will give an overview of an industrial pre-ranking system, then section 3 will introduce the details of COLD, including issues of model design, optimization of computing power cost and the whole infrastructure, section 4 and 5 will give experimental comparison and conclusion.

\section{Overview of pre-ranking system}

As illustrated in figure \ref{intro:cascade}, pre-ranking can be viewed as a connecting link between matching and ranking modules. It receives the result of matching and performs a rough selection to reduce the size of the candidate set for the following ranking module. Take the display advertising system in Alibaba as an example, the size M of the candidate set that is fed into the pre-ranking system often reaches ten thousand. Then the pre-ranking model selects top N candidates by certain metrics, e.g. eCPM  (expected Cost Per Mille) for advertising system \footnote{eCPM = pCTR * bid for CPC (Cost Per Click) bidding based advertising system}. The magnitude of N is usually several hundred. These winning N candidates are further ranked by a complex ranking model to get the final results to be displayed to users. 

Generally speaking, pre-ranking shares similar functionality of ranking. The biggest difference lies in the scale of the problem. Obviously, the size of candidates to be ranked is 10x or larger for the pre-ranking system than the ranking system. Directly apply ranking models in the pre-ranking system seems impossible, which will face the great challenge of computing power cost. How to balance the model performance and the computing power it costs is the key consideration for designing the pre-ranking system.        

\subsection{Vector-Product based DNN Model}
Driven by the success of deep learning, vector-product based DNN model \cite{youtube:recommend} has been widely used in pre-ranking systems and achieves state-of-the-art performance. As shown in Figure \ref{cmp},  architecture of vector-product based DNN model consists of two parallel sub neural networks. User features are fed to the left sub network and ad features to the right.  For each sub network, features are fed into the embedding layer first and then concatenated together, followed by FC (Fully Connected) layers. In this way, we obtain two fix-size vectors $\bm{v_u}$ and $\bm{v_a}$ which represents the user and ad information respectively. Finally, the pre-ranking score $p$ is calculated as follows:

\begin{equation}
 p = \sigma(\bm{v_u}^T \bm{v_a}), ~~~ where ~~~ \sigma(x)=\frac{1}{1+e^{-x}}.
\end{equation}
Training of the vector-product based DNN model follows the same way as the traditional ranking model. In order to focus on the key part of the pre-ranking model, we omit the details of training. Readers can refer to previous work such as   
\cite{zhou2018deep,zhou2019dien} for detailed introduction. 

\begin{figure}[th]
    \begin{center}  
      \includegraphics[width=0.45\textwidth] {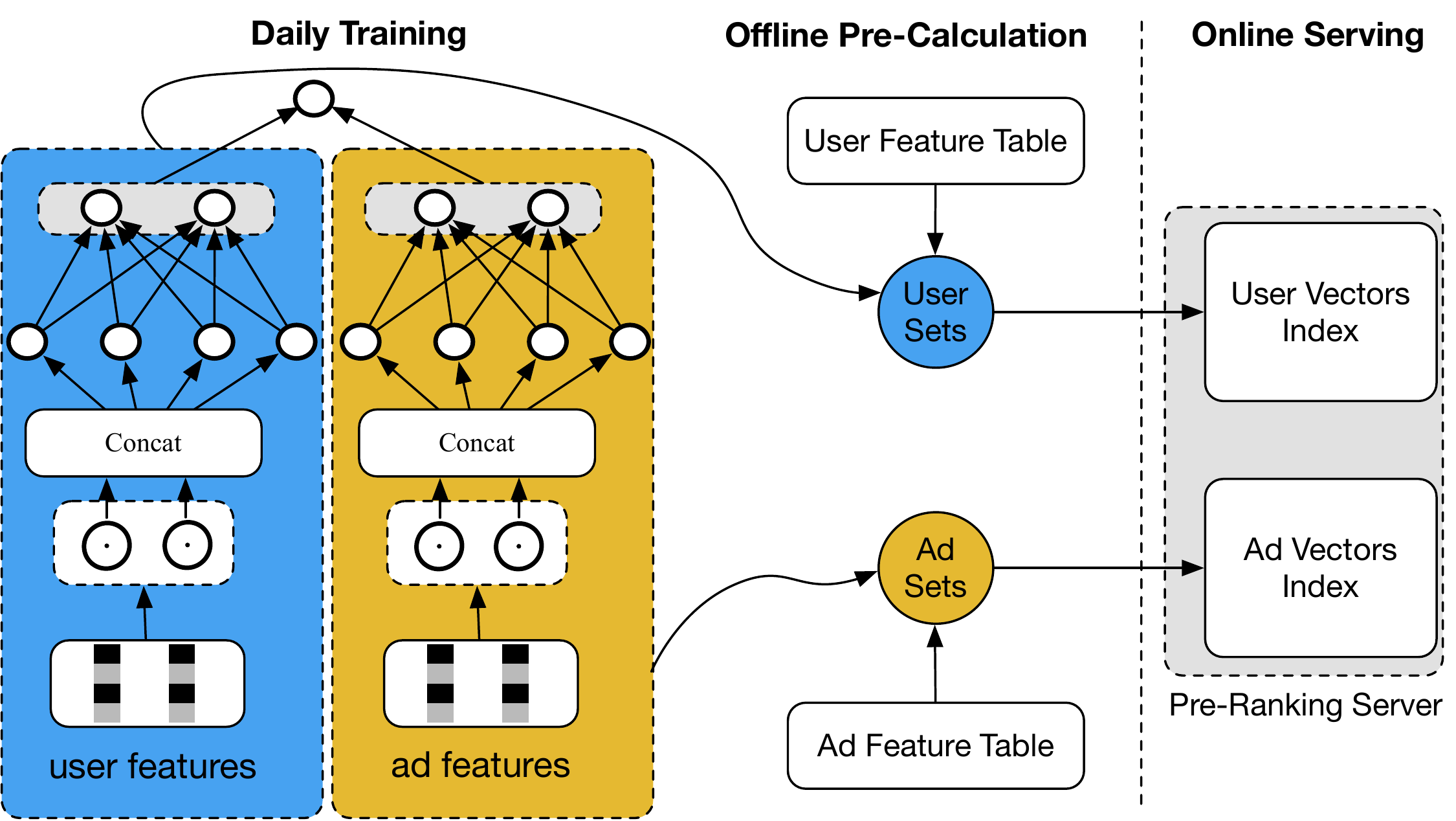} \\
      \caption{Infrastructure of pre-ranking system with vector-product based DNN model.}
      \label{intro:infra-vector}
    \end{center}
  \end{figure}

\subsection{Shortcomings of the Pre-Ranking System with Vector-Product based DNN Model}
The vector-product based DNN model is efficient in latency and computing resources. Vectors of $\bm{v_u}$ and $\bm{v_a}$ can be pre-calculated separately in an offline manner and score $p$ can be calculated online. This makes it friendly enough for tackling the challenge of computing power costs. Figure \ref{intro:infra-vector} illustrates the classic implementation of infrastructure.  Compared with previous generations of the pre-ranking model, the vector-product based DNN model achieves significant performance improvement. 

However, the vector-product based DNN model pays too much attention to reduce the computing power cost, by restricting the model to be in the vector-product form, this results in suboptimal performance. We summarize the shortcomings as follows:  
\begin{itemize}
    \item The model expression ability is limited by the vector-product form, and can not utilize the user-ad cross features. Previous work \cite{Zhu:2018gi} has shown the obvious superiority of incorporating complex deep models over vector-product form networks. 
    \item The user and ad vectors of $\bm{v_u}$ and $\bm{v_a}$  need to be pre-calculated offline by enumerating all users and ads, to reduce the computing resources and optimize the latency.  For businesses with hundreds of millions of users and tens of millions of ads, the pre-calculation usually costs several hours, making it hard to be adapted to the data distribution shift. This will bring great hurt of model performance when the data changes dramatically (e.g., Double 11 event in China).  
    \item The model update frequency is also affected by the system implementation. For the vector-product based DNN model, the daily switch between versions of user/ad vectors indexes needs to be executed at the same time, which is hard to be met with two indexes stored in different online systems. To our experience, the delayed switch also hurts the model performance.     
\end{itemize}

These shortcomings of the pre-ranking system with vector-product based DNN model originates from the excessive pursuit of computing power reduction and are hard to be tackled completely. In the following sections, we will introduce our new solution, which breaks the classic designing methodology for the pre-ranking system.


\section{COLD: computing power cost-aware online and lightweight deep pre-ranking system}
In this section, we will introduce our newly designed pre-ranking system COLD in detail. The core idea behind COLD is to take into consideration of both model design and system design. Computing power cost in COLD is also a variable that can be optimized jointly with model performance. In other words, COLD is a flexible pre-ranking system and the trade-off between model performance and computing power cost is controllable. 

\begin{figure*}[pt]
    \begin{center}
\centering
      \includegraphics[width=0.8\textwidth] {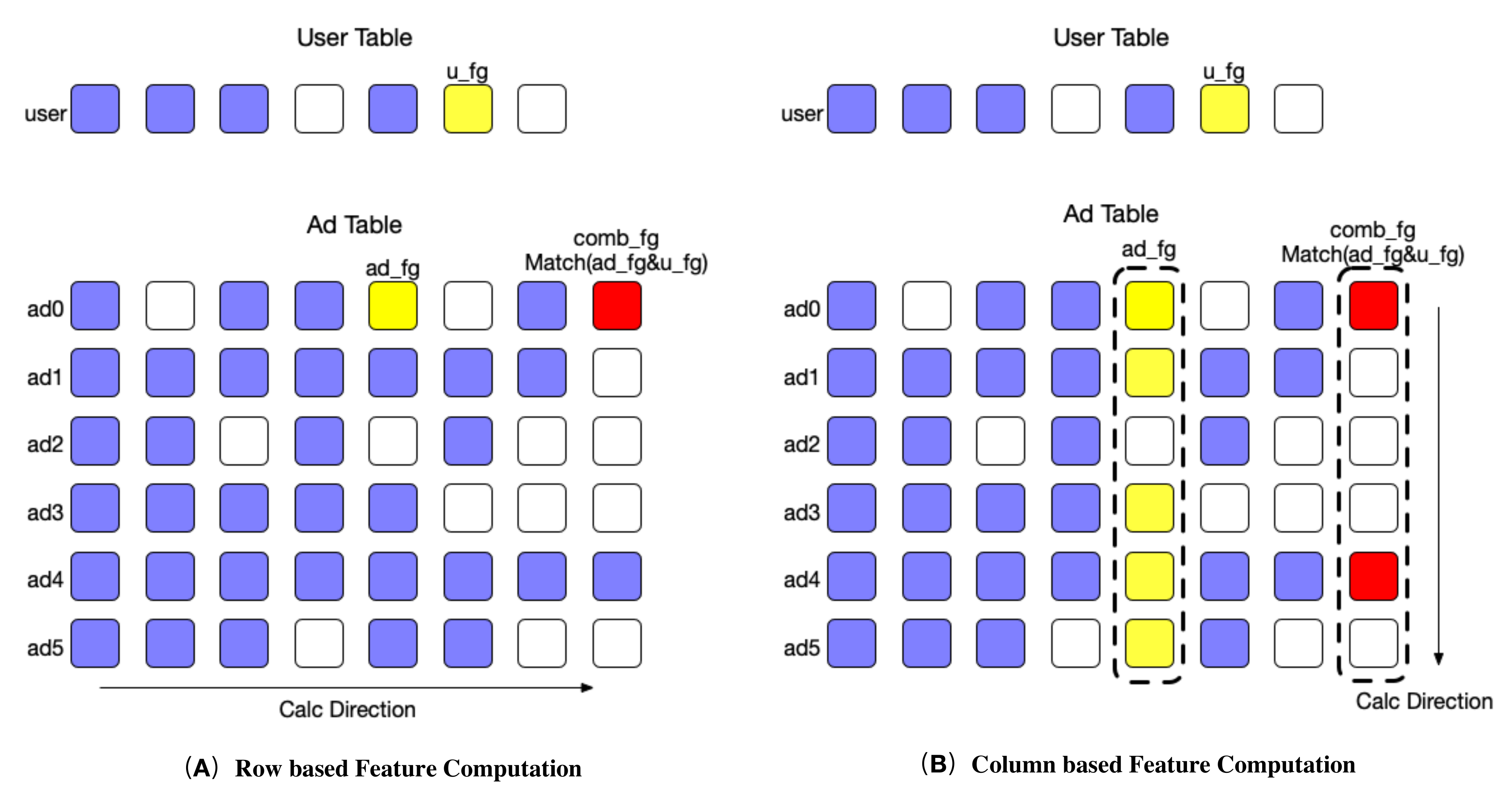} 
      \caption{Row V.S. Column based computations. Column based method is cache-friendly and enables further acceleration.} 
          \label{fc}
    \end{center}
  \end{figure*}

\subsection{Deep Pre-Ranking Model in COLD}

Unlike the vector-product based DNN model, which reduces computing power cost by restricting model architecture and thus causes loss of model performance, COLD allows applying arbitrary complex architecture of deep models to ensure the best model performance. In other words, SOTA deep ranking models can be used in COLD. For example, in our real system, we take \textbf{GwEN} (group-wise embedding network, referred as baseModel in \cite{zhou2018deep}) as our initial model architecture, which is an early version of the online model in our ranking system. Figure \ref{cmp} illustrates GwEN, which is a fully connected layer with the concatenation of feature group-wise embedding as inputs.  Note that cross features are also included in GwEN network.

Of course, the computing power cost of online inference by applying deep rank models with complex architecture directly is unacceptable, with a larger size of the candidate set to be ranked in the pre-ranking system. To tackle the critical challenge, we apply two ways of optimization strategy: one way is to design a flexible network architecture that can make a trade-off between model performance and computing power cost, the other way is to explicitly reduce computing power cost by applying engineered optimization tricks for inference acceleration.  

\subsection{Design of Flexible Network Architecture} 
Generally speaking, we need to introduce suitable designs of network architecture to get a lightweight version of the deep model from the full version of the initial GwEN model. Techniques such as network pruning, feature selection, and neural architecture search, etc. can be applied in this task. In our real practice, we choose the feature selection approach which is convenient for a controllable trade-off between model performance and computing power cost. Other techniques are also applicable, which we leave readers for further trying 
\cite{Huang:2019h,Liu:2020t}.

Specifically, we apply the SE (Squeeze-and-Excitation) block \cite{hu2018squeeze} for feature selection. SE block is firstly used in CV (Computer Vision) to explicitly model the inner-dependencies between channels. Here we use SE block to get the importance weights of group-wise features and select the most suitable ones in COLD, by measuring both model performance and computing power cost.  

\subsubsection*{\textbf{Importance weight calculation}}
 Let $e_i$ denote the embedding of $i_{th}$ input feature group. The total number of the feature groups is $M$.
The SE Block squeezes the input $e_i$ into a scalar value of weight $s_i$, which is calculated as:
\begin{equation}
s = \sigma(\bm{W}[e_1,...,e_m]+b),
\end{equation}
Where $\mathbf{s} \in \mathbb{R}^{M}$ is a vector,  $\mathbf{W} \in \mathbb{R}^{k \times M}, b \in \mathbb{R}^{M}$. $\mathbf{W}$ and $b$ are learn-able parameters. 
Then the new weighted embedding $v_i$ is calculated by field-wise multiplication between the embedding $e_i$ and the importance weight $s_i$.

\subsubsection*{\textbf{Feature group selection}}
The weight vector $\textbf{s}$ represents the importance of each feature group. We use the weight to rank all feature groups and select K groups of features with top weights. Then an offline test is conducted to evaluate the model performance and system performance of the candidate lightweight version of the model with selected  K groups of features. Metrics include  GAUC\cite{zhou2018deep}, QPS (Queries Per Seconds, which measures the throughput of the model), and RT (return time, which measures the latency of model). With several heuristic tries of number K, i.e., groups of features, we finally choose the version with the best GAUC  under a given constraint of system performance as our final model. 
In this way, the trade-off between model performance and computing power costs can be conducted in a flexible way.

\subsection{Engineered Optimization Tricks} 
Apart from reducing the computing power cost by flexible network architecture design, which is hard to avoid the hurt of model performance to some degree, we also apply various optimization tricks from an engineering view,  to further bring space for COLD to apply more complex deep models to reach better performance. 

Here we introduce the hands-on experience in the case of our display advertising system in Alibaba. Situations may vary from system to system. Readers can make choices according to the actual situation. In our display advertising system, the online inference engine of the pre-ranking module mainly contains two parts: feature computation and dense network computation. In feature computation, the engine pulls user and ad features from the indexing system and then computes cross-features. In dense network computation, the engine first turns features into embedding vectors and concatenate them as the input of the network.

\subsubsection*{\textbf{Parallelism at All Level}}
To achieve low latency and high throughput inference with low computing power cost, leveraging parallel computing is important. Our system, therefore, applies parallelism whenever is possible. Fortunately, the pre-rank score of different ads is independent of each other. This means they can be computed in parallel with the cost that there may be some duplicated computation related to user features. 

At a high level, one front-end user query will be split into several inference queries. Each query handles parts of the ads and the results will be merged after all the queries return. Therefore there are trade-offs when deciding how many queries to split. More queries mean few ads for each query and therefore lower latency for an individual query. But too many queries can also lead to huge duplicated computation and system overhead. Also, as the queries are implemented using RPC (Remote Procedure Call) in our system, more queries mean more network traffic and could have a higher chance of delay or failure. 

When handling each query, multi-thread processing is used for feature computation. Again, each thread handles parts of ads to reduce the latency. Finally, when executing dense network inference, we use GPU to accelerate the computation.

\subsubsection*{\textbf{Column based Computation}}
Traditionally, feature computation is done in a row based manner: ads are being processed one by one. However, such a row based method is not cache-friendly. Instead, we use a column-based method to put computations of one feature column together. Figure \ref{fc} illustrates the two kinds of computation modes. By doing so, we can use techniques like SIMD (Single Instruction Multiple Data) to accelerate feature computation.

\subsubsection*{\textbf{Low precision GPU calculation}}
For COLD model, most of the computation is the dense matrix multiplication, which leaves optimization space. In NVIDIA's Turing Architecture, the T4 GPU provides extreme performance for Float16 and Int8 matrix multiplication which perfectly fits our case. The theoretical peak FLOPS for Float16 can be 8 times higher than Float32. However, the Float16 loses some precision. In practice, we found that for some scenario, as we use sum-pooling for some feature groups, the input of the dense network could be a very large number and exceed Float16 representation. To solve this, one solution is to use normalization layers like the batch-norm layer. However, the BN layer itself contains moving-variance parameters whose scale could be even larger. This means the computation graph needs to be mix-precision \cite{Micikevicius:2017vd} that fully-connected layers use Float16 and batch-norm layers use Float32. Another approach is to use a parameter-free normalization layer. For example, the logarithmic function can easily transform large numbers into a reasonable range. However, log() function can not handle negative values and can result in a huge number when input is near zero. Therefore, we design a piece-wised smooth function called linear-log operator to handle that unwanted behavior, as shown in Eq. \ref{eq:log}

\begin{equation}
    linear\_log(x) = \left\{
        \begin{array}{rcl}
        &-log(-x)-1   & {x< -1} \\
        &x   &  {-1\leq x\leq 1}.\\
        &log(x)+1   & {x>1}
        \end{array}
    \right.
\label{eq:log}
\end{equation}

\begin{figure}
     \begin{center}
       \includegraphics[width=0.45\textwidth] {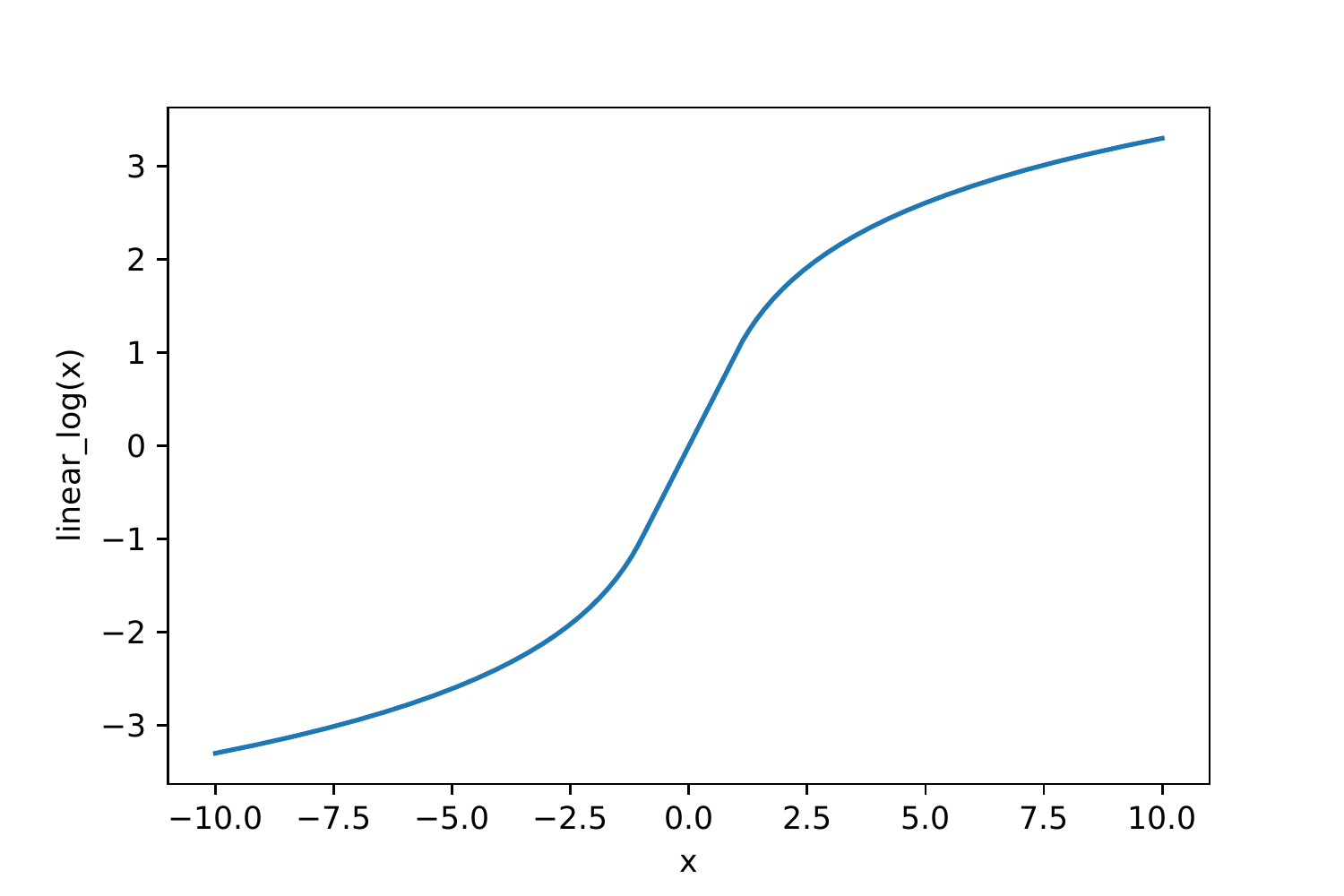} \\
       \caption{\label{fig:log}The linear\_log function}
     \end{center}
\end{figure}

The graphics of the linear\_log() function can be seen in Figure \ref{fig:log}. It transforms Float32 numbers into a reasonable range. So if we put a linear\_log operator in the first layer, it guarantees the input of the network would be small. Also, the linear\_log() function is $C^1$ continuous, so that it won't make network training harder. In practice, we found that after adding this layer, the network can still reach the same accuracy comparing to the original COLD model.

After using Float16 for inference, we found that the running time of CUDA kernel drops dramatically, and the kernel launching time becomes the bottleneck. To boost actual Querys Per Seconds (QPS), we further use MPS (Multi-Process Service) to reduce overhead when launching the kernels. Combining Float16 and MPS, the engine throughput is twice as before.

\begin{figure}[th]
    \begin{center}  
      \includegraphics[width=0.45\textwidth] {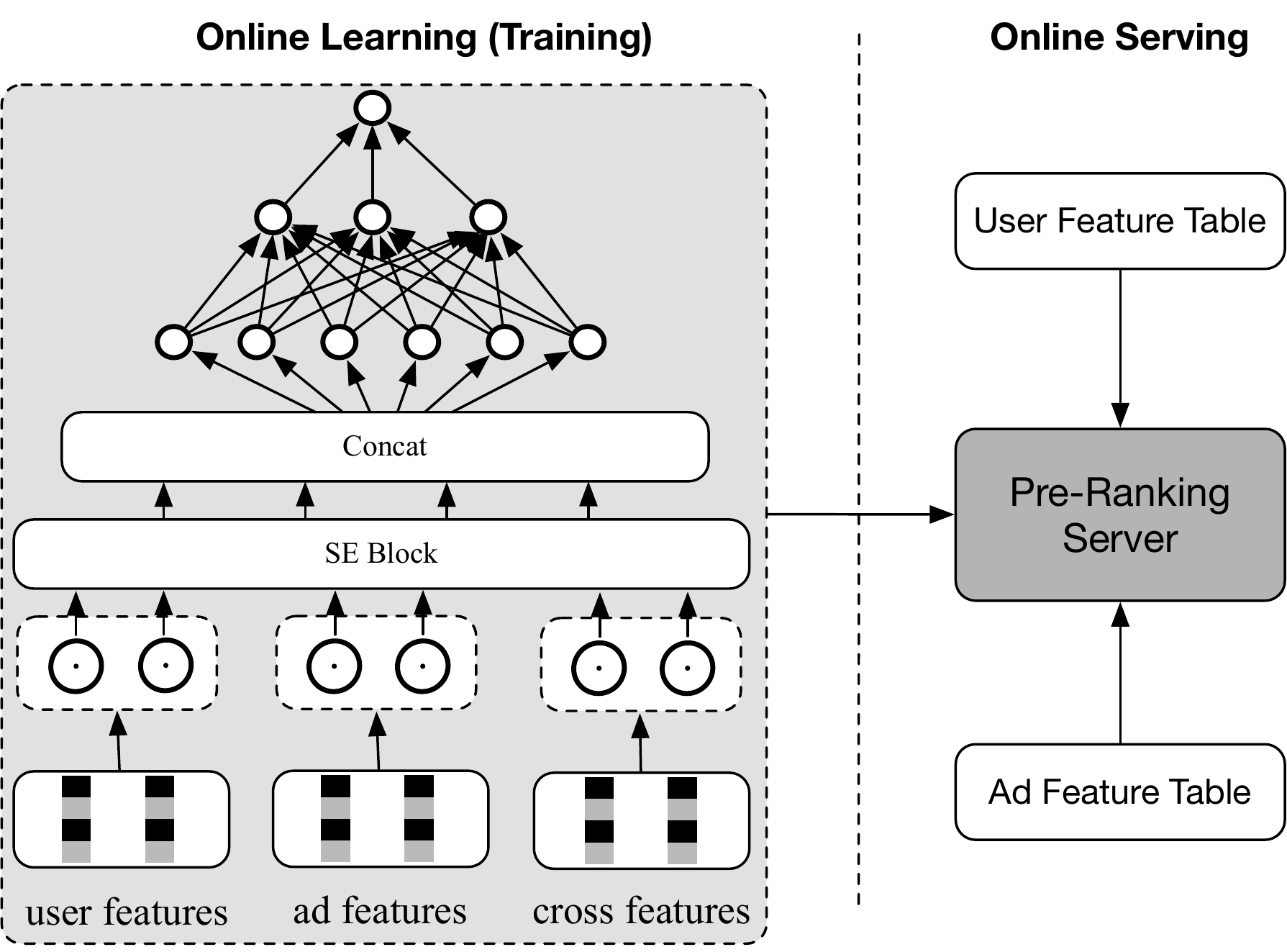} \\
      \caption{Infrastructure of fully online infrastructure of COLD pre-ranking system.}
      \label{intro:infra-cold}
    \end{center}
  \end{figure}

\subsection{Fully Online Infrastructure of COLD Pre-Ranking System} 
Benefiting from the unrestricted model architecture, COLD can be implemented under a fully online infrastructure: both training and serving are executed in an online manner, as illustrated in Figure \ref{intro:infra-cold}. From the industry view, it's the best system practice currently that the ranking system owns. This infrastructure benefits COLD in two folds:   
\begin{itemize}
    \item Online learning of COLD model brings its excellent ability to handle the challenge of data distribution shift. To our experience, as shown in the next experimental section, the improvement of model performance that COLD model over vector-product based DNN model is more significant when the data changes dramatically (e.g., Double 11 event in China). Besides, COLD model is more friendly to the new ads with online learning.  
    \item The fully online pre-ranking system of COLD provides us with a flexible infrastructure that supports efficient new model developing and online A/B testing. Remember that for vector-product based DNN model, vectors of the user and ad side need to be pre-calculated offline and load to inference engine by index. Thus it involves development over several systems to conduct A/B testing of two versions of vector-product based DNN model. To our experience, the typical time cost to get a solid A/B testing result is several days, which in turn is several hours for COLD. Besides, fully online serving also helps COLD to avoid delayed switch that the vector-product based DNN model suffers.    
\end{itemize}


\section{Experiments}

We conduct careful comparisons to evaluate the performance of the proposed pre-ranking system COLD. As an industrial system, comparisons are conducted on both model performance and system performance. To the best of our knowledge, there are merely public datasets or pre-ranking systems for this task. The following experiments are executed in the online display advertising system in Alibaba.

\subsection{Experimental Settings}
The strongest baseline of COLD model is the SOTA vector-product based DNN model, which is the latest version of the online pre-ranking model in our display advertising system. 

Both COLD model and vector-product based DNN model are trained with more than 90 billion samples, which are collected from logs of the real system. Note that the vector-product based DNN model shares the same user and ad features with COLD model. The vector-product based DNN model could not introduce any user-ad cross features, while COLD model uses user-ad cross features. For a fair comparison, we also evaluate the performance of COLD model with different groups of cross features.  
For COLD model, the feature embedding vectors are concatenated together and then fed into a fully connected network~(FCN). The structure of this FCN is $D_{in} \times 1024 \times 512 \times 256 \times 128 \times 64 \times 2$, where $D_{in}$ means the dimension of the concatenated embedding vectors of selected features. For the vector-product based model, the FC layers are set by $200 \times 200 \times 10$. The dimension of input feature embedding is set to be 16 for both two kinds of models. We use Adam solver to update the model parameters. GAUC \cite{zhou2018deep} is used as the metric to evaluate the offline performance of the models. Besides, we introduce a new metric of top-k recall, to measure the alignment degree between the pre-ranking model and subsequent ranking model. The top-k recall rate is defined as:
\begin{equation}
  recall = \frac{|\{\text{top k ad candidates}\} \cap \{ \text{top m ad candidates}\}|}{|\{\text{top m ad candidates}\}\}|},
\end{equation}
where the $\text{top k candidates}$ and the $\text{top m candidates}$ are generated from the same candidate set, which is the input of the pre-ranking module. The top k ad candidates are ranked by the pre-ranking model and the top m ad candidates are ranked by ranking model. The ranking metric is eCPM (expected Cost Per Mille, eCPM = pCTR * bid). 
In our experiments, the ranking model uses DIEN \cite{zhou2019dien}, a previous version of the online ranking system. 

For evaluation of system performance, we use metrics including QPS (Queries Per Seconds, which measures the throughput of the model), RT (return time, which measures the latency of model). These metrics reflect the computing power cost of the model under the same size of the candidate set to be pre-ranked. Roughly speaking, larger QPS under lower RT means lower computing power cost for a given model.

\subsection{Evaluation on Model Performance}
Table \ref{exp:offline} shows the offline evaluation results of different models. 
We can see that COLD maintain a comparable GAUC with our previous version ranking model DIEN, and achieves significant improvement both in GAUC and Recall compared with the vector-product based model.

\begin{table}[t]
  \caption{Offline evaluation results}
\label{exp:offline}    
  \begin{tabular}{lcl}
    \toprule
     Method & GAUC & Recall\\
    \midrule
        Vector-Product based DNN Model & 0.6232 & 88\% \\ 
        COLD  & 0.6391 & 96\%\\ 
        DIEN  & 0.6511 & 100\%\\ 
  \bottomrule
\end{tabular}
\end{table}

We also conduct careful online A/B testing. Table \ref{exp:online} shows the lift of COLD model over the vector-product based DNN model. In normal days, COLD model achieves 6.1\%\ CTR and 6.5\%\ RPM (Revenue Per Mille) improvement, which is significant to our business. Moreover, the improvement turns to be 9.1\%\ CTR and 10.8\%\ RPM during the double 11 event. It proves the value of the fully online infrastructure of COLD which can enable the model to be adapted to the newest data distribution when the data changes dramatically.   
  
\begin{table}
  \caption{Results of online A/B testing by comparing COLD model with vector-product based DNN model.}
\label{exp:online}    
  \begin{tabular}{lcl}
    \toprule
     Time & CTR lift & RPM lift \\ 
    \midrule
     Normal Days & +6.1\%\ & +6.5\%\ \\ 
     Double 11 Event& +9.1\%\ & +10.8\%\ \\ 
  \bottomrule
\end{tabular}
\end{table}

\subsection{Evaluation on System Performance}

\begin{table}
  \caption{Comparison of system performance of pre-ranking system that servers with different models}
\label{tab:sys}
  \begin{tabular}{lcc}
    \toprule
       Model & QPS & RT  \\ 
    \midrule
       Vector-Product based DNN Model  & 60000+ & 2ms \\ 
       COLD  & 6700 & 9.3ms  \\ 
       DIEN & 629 & 16.9ms \\ 
  \bottomrule
\end{tabular}
\end{table}

We evaluate the QPS and RT of the pre-ranking system that serves with different models. Table \ref{tab:sys} gives the results. Vector-product based model runs on a CPU machine with 2 Intel(R) Xeon(R) Platinum 8163 CPU @ 2.50GHz (96 cores) and 512GB RAM. COLD models and DIEN run on a GPU machine with NVIDIA T4 equipped. In this time, the vector-product based DNN model achieves the best system performance, which is as expected. The computing power of DIEN costs the most. COLD achieves the balance.  


\subsection{Ablation Study of COLD}
To further understand the performance of COLD, we conduct experiments on both model design view and engineered optimization techniques view. 
For the latter one, as it is hard to decouple all the optimization techniques from the integrated system and compare each of them, here we conduct an evaluation on the most significant factor of low precision computation of GPU.  

\begin{table}
\begin{threeparttable}
  \caption{Trade-off performance of pre-ranking system with different versions of COLD model}
\label{tab:tradeoff}
  \begin{tabular}{lccc}
    \toprule
       Model & QPS & RT & GAUC \\ 
    \midrule
       COLD (No Cross Features)& 6860 & 8.6ms & 0.6281 \\ 
       \textbf{COLD} \tnote{*} & 6700 & 9.3ms & 0.6391 \\        
       COLD (All Features)& 2570 & 10.6ms & 0.6467\\ 
  \bottomrule
\end{tabular}
\begin{tablenotes}
\item[*] This is the balanced version of COLD model that we use as product version. It users partial of cross-features. 
\end{tablenotes}
\end{threeparttable}
\end{table}

\textbf{Trade-off performance of the pre-ranking system with different versions of COLD model}.
In the model design stage, we use the SE block to get the feature important weights and select different groups of features as the candidate version of models. Then we conduct the offline experiments to evaluate the models with QPS, RT, and GAUC.  Table \ref{tab:tradeoff} shows results. Obviously, computing power cost of COLD model varies with different features. This fits our design of flexible network architecture. COLD model with more cross features achieves better performance, which also increases the burdens for online serving correspondingly. In this way, we can make a trade-off of model performance and computing power cost. In our real system, we choose the balanced version manually by experience.

\textbf{Comparison of computing with different GPU precisions.} The experiments run on a GPU machine with NVIDIA T4. When running the experiments, we gradually increase the QPS from the client-side, until the more than 1\% of the server responding time begins to exceed the latency limit. We then record the current QPS as usable QPS. As shown in Table \ref{tab:qps_cuda}, the Float32 version has the lowest usable QPS. Using Float16 alone can improve about 21\% of the usable QPS. Combining Float16 and CUDA MPS, we can double the usable QPS comparing to Float32, and can fully utilize the GPU without exceeding the latency limit.

\begin{table}
  \caption{Comparison of system performance by computing with different GPU precisions}
\label{tab:qps_cuda}
  \begin{tabular}{cc}
    \toprule
      Precision & QPS  \\ 
    \midrule
      COLD (Float32) & 2800  \\ 
      COLD (Float16) & 3400 \\ 
      COLD (Float16+MPS) & 6700 \\ 
  \bottomrule
\end{tabular}
\end{table}



\section{Conclusion}

In this paper, we introduce our new generation of pre-ranking system COLD in detail . It is designed from a brand-new perspective. Instead of saving computing power with hard restriction of model architecture which causes loss of model performance, COLD takes into consideration both model design and system design. Computing power cost in COLD is also a variable that can be optimized jointly with model performance. With the co-design of the model architecture and computing power cost, COLD turns to be a flexible pre-ranking system that the trade-off between model performance and computing power cost is controllable. This new pre-ranking system enables a better pursuit of model performance. Experiments show COLD model achieves more than 6\%\ RPM lift over vector-product based DNN model, our latest online version of the pre-ranking model, which is significant to the business. Besides, COLD can be implemented with a fully online infrastructure for both training and serving, achieving the best system practice that the current ranking system owns. Since 2019, COLD has been deployed in the display advertising system in Alibaba and serves the main traffic of almost all products, contributing a significant business revenue growth.

\bibliographystyle{ACM-Reference-Format}
\bibliography{dqm}

\end{document}